\documentclass[epj]{svjour}  
 \usepackage{graphics}
 
 \def\T{\textstyle}
 \def\l{\left}
 \def\r{\right}
 \def\nf{n_{\!f}}
 \def\be{\begin{equation}}
 \def\ee{\end{equation}}
 \def\bea{\begin{eqnarray}}
 \def\eea{\end{eqnarray}}
 \def\ksim{\mathrel{\rlap{\lower0.2em\hbox{$\sim$}}\raise0.2em\hbox{$<$}}}
\begin{document}

 \title{Rethinking the QCD collisional energy loss}
 \author{A.~Peshier}
 \institute{Institut f\"{u}r Theoretische Physik,
 	Universit\"{a}t Giessen, D-35392 Giessen, Germany}
 \date{Received: date / Revised version: date}

\abstract{
It is shown that to leading order the collisional energy loss of an energetic  parton in the hot quark gluon plasma reads $dE/dx \sim \alpha(m_D^2)T^2$, where the scale of the coupling is determined by the (parametrically soft) Debye screening mass.
Compared to previous expressions derived by Bjorken and other authors, $dE^B/dx \sim \alpha^2 T^2 \ln(ET/m_D^2)$, the rectified result takes into account the running of the coupling, as dictated by quantum corrections beyond tree level. As one significant consequence, due to asymptotic freedom, the QCD collisional energy loss becomes independent of the jet energy in the limit $E \gg T$.
It is advocated that this resummation improved perturbative result might be useful to (re-)estimate the collisional energy loss for temperatures relevant in heavy ion phenomenology.
}

\maketitle

\section{Introduction} \label{intro}
The quenching of jets observed at RHIC \cite{jet quenching EXP} has become one of the key arguments for the creation of a `new state of matter' in heavy ion collisions.
Being, {\em inter alia}, attributable to the energy loss of energetic partons (`jets'), it indeed allows to infer specific properties of the traversed matter.
In a quark gluon plasma (QGP) there are two different loss mechanisms: elastic collisions with deconfined partons \cite{Bjorken:1982tu,Braaten:1991jj}, or induced gluon radiation \cite{Wang:1991xy,Baier:1996kr,Zakharov:1996fv,Gyulassy:2000fs}.
Presuming a dominance of the second effect, experimental findings have often been interpreted in terms of a purely radiative energy loss. However, the data-adjusted model parameters (either $\hat q$ or $dN_g/dy$, depending on the approach) seem debatable because they are considerably larger than theoretically expected or even in conflict with a strong constraint from $dS/dy$, see e.\,g.\ \cite{Muller:2005wi}.

An obvious remedy for this situation is a sizable collisional contribution to the energy loss, as considered recently in \cite{Mustafa:2003vh,Wicks:2005gt}. These considerations as well as all previous studies were based on formulae of the generic form $dE^B/dx \sim \alpha^2 T^2 \ln(ET/m_D^2)$ due to Bjorken, which can be considered an adaption of the (QED) Bethe-Bloch formula. Evidently, such estimates depend crucially on the size of the coupling $\alpha$. For practical estimates it has been common practice to {\em choose} a number for $\alpha$ by evaluating the running coupling at a `typical' thermal momentum scale, often of the order of $2\pi T$.
This {\em ad hoc} `prescription' was scrutinized only recently \cite{Peshier:2006hi}.

In fact, Bjorken's formula is based on a tree level scattering amplitude -- hence the parameter $\alpha$ in $dE^B/dx$ is to be considered a {\em fixed} number. This level of approximation is acceptable in the QED analog, where the strength of the electromagnetic coupling is virtually constant over a huge range of momenta. The strong coupling, on the other hand, varies considerably in a momentum range which can be probed by the energy loss in heavy ion collisions. Thus, in QCD, quantum corrections to the tree level scattering amplitude, which induce the running of the coupling, have to be taken into account on conceptual grounds.

Before discussing the implications of this statement for the collisional energy loss, I will briefly review the derivation of Bjorken's formula.

\section{Bjorken's formula}
Bjorken \cite{Bjorken:1982tu} considered the propagation of a jet $j$ through a static QGP at a temperature $T \gg \Lambda$, where the coupling is small and a perturbative approach is appropriate.
Its mean energy loss per length can be calculated from the rate of binary collisions with partons of the medium, as determined by the flux and the cross section,
\be
	\frac{dE_j}{dx}
	=
	\sum_s
	\int_{k^3} \rho_s(k)\, \Phi 
	\int dt\, \frac{d\sigma_{\!js}}{dt}\, \omega \, .
\label{EQ.dEdx_start}
\ee
Here $\rho_s = d_s n_s$ is the density of scatterers $s$, with $d_g = 16$ and $d_q = 12\nf$ being the gluon and quark degeneracies for $\nf$ light flavors, and $n_s \rightarrow n_\pm^{\rm id}(k) = \l[ \exp(k/T) \pm 1 \r]^{-1}$ in the ideal gas approximation. Furthermore, $\Phi$ denotes a dimensionless flux factor, $t$ the 4-momentum transfer squared, and $\omega = E-E'$ the energy difference of the incoming and outgoing jet.
Scatterings with small momentum transfer dominate, thus the cross section can be approximated by the $t$-channel contribution,
\be
	\frac{d\sigma_{\!js}}{dt}
	=
	2\pi C_{\!js}\, \frac{\alpha^2}{t^2} \, ,
\label{EQ.dsigmadt}
\ee
with $C_{qq} = \frac49$, $C_{qg} = 1$, and $C_{gg} = \frac94$.
For $E$ and $E'$ much larger than the typical momentum of the thermal scatterers, $k \sim T$, the relation of $t$ to the angle $\theta$ between the jet and the scatterer simplifies to
\be
	t = -2(1-\cos\theta)k\omega \, ,
\label{EQ.t(eps)}
\ee
and the flux factor can be approximated by $\Phi = 1-\cos\theta$.

At this point, Bjorken integrated in Eq.~(\ref{EQ.dEdx_start})
\be
	\Phi\int_{t_1}^{t_2}\! dt\, \frac{d\sigma_{\!js}}{dt}\, \omega
	=
	\frac{\pi C_{\!js} \alpha^2}{-k}
	\int_{t_1}^{t_2} \frac{dt}t
	=
	\frac{\pi C_{\!js} \alpha^2}{k}\,
	\ln\frac{t_1}{t_2} \, ,
\label{EQ.Bjorkens_t-int}
\ee
imposing both an IR and UV regularization. 
The soft cut-off is related to the Debye mass, $|t_2| = \mu^2 \sim m_D^2 \sim \alpha T^2$, describing the screening of the exchanged gluon by the medium.
The upper bound of $|t|$ was reasoned to be given by the maximum energy transfer: very hard transfers, say $\omega \approx E$, do not contribute to the energy loss; in this case the energy is collinearly relocated to the scatterer.
Assuming $\omega_{max} = E/2$ implies $t_1 = -(1-\cos\theta)kE$, thus
$dE_j^B/dx = \pi\alpha^2 \sum_s C_{\!js}\int_{k^3} k^{-1} \rho_s \ln\!\l((1-\cos\theta)kE/\mu^2\r)$.
Replacing now, somewhat pragmatically, the logarithm by $\ln(2\langle k \rangle E/\mu^2)$ and setting $\langle k \rangle \rightarrow 2T$, Bjorken arrived at
\be
	\frac{dE_{q,g}^B}{dx}
	=
	\l( \T\frac23 \r)^{\!\pm 1}
	2\pi \l( 1+\T\frac16 \nf \r) \alpha^2 T^2 \ln\frac{4TE}{\mu^2} \, ,
\label{EQ.dEdx_Bjorken}
\ee
which differs for quarks and gluons only by the prefactor.

Various refinements of Bjorken's illustrative calculation, as e.\,g.\ \cite{Braaten:1991jj}, aim at the precise determination of the cut-offs. The generic form of these results agrees with Eq.~(\ref{EQ.dEdx_Bjorken}), with $\mu \rightarrow m_D$ in the logarithm and the factor 4 replaced by some function of $\nf$.

\section{QCD collisional loss with running coupling}
As already motivated in the Introduction, the running of the coupling should be regarded when its strength varies notably in the range of probed momenta. In the present case of a $t$-channel dominated observable, the relevant scale is the virtuality of the exchanged gluon, $t \in [t_1,t_2]$. For a representative situation ($T = 0.3\,$GeV, $E = 10\,$GeV), one roughly finds $|t_1| \sim 10\,$GeV$^2$ and $|t_2| \sim 1\,$GeV$^2$, thus for phenomenological estimates it is conspicuously essential to take into account the running of the coupling.

Formally, the running coupling arises in the course of renormalization of the (divergent) loop corrections to the tree level scattering amplitude. Referring to \cite{Peshier:2006hi} for some details of the elementary considerations, the resulting resummation {\em improved} cross section is basically the expression (\ref{EQ.dsigmadt}) with the (constant) parameter $\alpha$ replaced by
\be
  \alpha(t) = \frac{b_0^{-1}}{\ln(|t|/\Lambda^2)} \, ,
  \label{EQ.alpha(t)}
\ee
where $\Lambda$ is the QCD parameter, and $b_0 = (11 - \frac23\, \nf)/(4\pi)$. Moreover, an infrared cut-off $\mu^2 \sim m_D^2$ (stemming from the thermal contributions to the gluon selfenergy) is to be understood implicitly, as already employed by Bjorken. 

At this point it should be noted that previous calculations leading to results of the form (\ref{EQ.dEdx_Bjorken}) are, strictly speaking, inconsistent. On the one hand, an IR cut-off, which arises only from loop corrections to the tree level approximation, is needed in the calculation to screen the long-range interaction in the medium. On the other hand side, the (renormalized) vacuum parts of the same loop corrections (which make the coupling run) were omitted.
This circumstance is the heart of the ambiguity that one has to {\em guess} a value of the parameter $\alpha$ in Eq.~(\ref{EQ.dEdx_Bjorken}) -- whereas a consistent calculation will actually specify the value of the coupling.

Taking into account the running coupling (\ref{EQ.alpha(t)}) alters completely the structure of the integral (\ref{EQ.Bjorkens_t-int}),\footnote{Aiming here
	only at a leading order result, details related to the broken Lorentz 
	invariance in the presence of a medium can be omitted.}
\bea 
	&&\Phi\int_{t_1}^{t_2}\! dt\, \frac{d\sigma_{\!js}}{dt}\, \omega
	\,=\,
	-\frac{\pi C_{\!js}}{k\, b_0^2}
	\int_{t_1}^{t_2} \frac{dt}{t\ln^2(|t|/\Lambda^2)}
	\nonumber \\
	&&=\,
	\frac{\pi C_{\!js}}{k\, b_0^2}
	\l. \frac1{\ln(|t|/\Lambda^2)} \r|_{t_1}^{t_2}
	\,=\,
	\frac{\pi C_{\!js}}{k\, b_0}
	\l[ \alpha(\mu^2) - \alpha(t_1) \r] . \rule{7mm}{0mm}
	\label{EQ.t-int}
\eea
Opposed to the former expression (\ref{EQ.Bjorkens_t-int}), the weighted cross section is manifestly UV finite -- due to the asymptotic weakening of the strong interaction, large-$t$ contributions are suppressed. For very hard jets, the integral (\ref{EQ.t-int}) becomes independent of the energy $E$, i.\,e., it is then controlled solely by the coupling at the screening scale. The necessary condition $|t_1| \sim TE \gg \mu^2 \sim m_D^2 \sim \alpha T^2$, is, for weak coupling, actually much less restrictive than the previous assumption $E \gg T$ to simplify the kinematics. 
Aspiring an extrapolation of the final result for $dE/dx$ to larger coupling, where the hierarchy of scales $\sqrt\alpha T \ll T$ breaks down, it is mentioned that the Debye mass is constrained by $m_D \ksim 3T$ (see, e.\,g., \cite{Peshier:2006ah}). Hence from this perspective, the collisional loss could become $E$-independent already for jet energies exceeding several GeV's -- provided, of course, that the present framework gives at least a semi-quantitative guidance at larger coupling, which will be advocated below.

As an aside, it is instructive to illustrate in more detail the relation of Eqs.~(\ref{EQ.Bjorkens_t-int}) and (\ref{EQ.t-int}). The latter (full) result has the structure
\[
	\int_a^b d\tau\, \frac1{\tau \ln^2 \tau}
	=
	\frac{\ln b - \ln a}{\ln a \ln b} \, ,
\]
which might be written in a `factorized' form,
\[
	\frac{\ln b - \ln a}{\ln^2 \tau^\star}
	=
	\frac1{\ln^2 \tau^\star}\int_a^b d\tau\, \frac1\tau \, ,
\]
where $\ln^2 \tau^\star = \ln a \ln b$. With the notation $a = \gamma^2 b$, one can, if applicable, approximate $\ln^2 \tau^\star = (\ln b + 2\ln\gamma) \ln b \approx (\ln b + \ln\gamma)^2$, hence $\tau^\star \approx \gamma b = \sqrt{ab}$.
Identifying $b \sim TE/\Lambda^2$ and $a \sim \mu^2/\Lambda^2$, one observes that Bjorken's `factorized' formula can be warrant -- provided $\alpha$ is replaced by the running coupling $\alpha(t)$ at an {\em intermediate} scale $t^\star \sim \sqrt{\mu^2 TE} \sim gT\sqrt{TE}$ (which then leads to a cancellation with the logarithm in Eq.~(\ref{EQ.Bjorkens_t-int})).

Returning to the full expression (\ref{EQ.t-int}), which is actually both simpler and physically more intuitive, the collisional energy loss of a parton approaches 
\be
	\frac{dE_j}{dx}
	=
	\pi \frac{\alpha(\mu^2)}{b_0} \sum_s C_{\!js}\int_{k^3} 
	\frac{\rho_s(k)}k
\label{EQ.dEdx_before_k-int}
\ee
in the infinite-energy limit.
With the ideal gas densities, the remaining integration yields a simple formula,
\be
	\frac{dE_{q,g}}{dx}
	=
	\l( \T\frac23 \r)^{\!\pm 1}
 	2\pi \l( 1+\T\frac16 \nf \r) \frac{\alpha(\mu^2)}{b_0}\, T^2 \, ,
 \label{EQ.dEdx}
\ee
which differs clearly from previous expressions as Bjorken's formula (\ref{EQ.dEdx_Bjorken}). Aside from the modified cut-off dependence discussed above, the QCD collisional loss is proportional to $\alpha$ (instead of $\alpha^2$). It is emphasized that the present considerations also show that the relevant scale for the coupling is the (parametrically soft) screening mass rather than a `characteristic' thermal (hard) scale $\sim T$, as commonly presumed.

In order to quantitatively compare the new result to previous estimates it is necessary to specify parameters, namely $\Lambda$ in Eq.~(\ref{EQ.alpha(t)}) and the cut-off $\mu$, i.\,e., the Debye mass. 
Similar renormalization arguments as above led in \cite{Peshier:2006ah} to the conclusion that the familiar leading-order expression $m_D^2 = \T\l( 1+\frac16 \nf \r) 4\pi\alpha T^2$ should be replaced by an {\em implicit} equation; to leading order 
\be
	m_D^2 = \T\l( 1+\frac16 \nf \r) 4\pi\alpha(m_D^2)\, T^2 \, ,
	\label{EQ.mD}
\ee
whose solution can be given in terms of Lambert's function.
Obviously, this {\em improved} perturbative formula is justified strictly only at temperatures $T \gg \Lambda$. Notwithstanding this, it is found in {\em quantitative} agreement with non-perturbative lattice QCD calculations down to $\tilde T \approx 1.2T_c$.\footnote{In contrast, the `conventional' expression,
	again with the common {\em choice} $\alpha \rightarrow \alpha(2\pi T)$, 
	deviates from the lattice results by an almost constant factor of 1.5 in a
	large temperature range, which is not much improved by the next-to-leading 
	order correction.}
I underline that the only `free' parameter, fitted as $\Lambda = 0.2\,$GeV, is not only right in the expected ballpark (refuting the possibility of a mere `coincidental' fit), but it also describes further non-perturbative results. As a matter of fact, the distance dependence of the strong coupling, as calculated from the heavy quark potential at $T = 0$ in lattice QCD, is also reproduced {\em quantitatively} by the elementary 1-loop formula (\ref{EQ.alpha(t)}) with $t \rightarrow 1/r^2$ -- most notably even for large distances corresponding to $\alpha(r) \approx 1$ \cite{Peshier:2006ah}.

It might be unexpected that {\em resummed} perturbation theory works beyond its expected range of validity\footnote{As argued e.\,g.\ in 
	\cite{Peshier:1998ei}, perturbative expansions could (at best) be asymptotic 
	series, implying an optimal truncation order {\em inversely} proportional 
	to the coupling. Hence for strong coupling, a leading order result might 
	indeed be more adequate than sophisticated high-order calculations  -- 
	somewhat opposed to naive intuition.}, 
but it is actually less surprising to find it appropriate {\em jointly} for both quantities. In fact, the vacuum potential and screening at $T>0$ are closely related: they are both described by the (renormalized) $t$-channel scattering amplitude \ldots which also determines the collisional energy loss.
Put differently: one can renormalize the theory at $T=0$ (i.\,e., determine once and for all $\Lambda$ from $V_{qq}(r)$), verify the applicability of the resummed perturbative approach for larger couplings as relevant near $T_c$ by successfully calculating $m_D(T)$, and give then {\em justified} estimates for $dE/dx$.

\begin{figure}
\resizebox{0.5\textwidth}{!}{\includegraphics{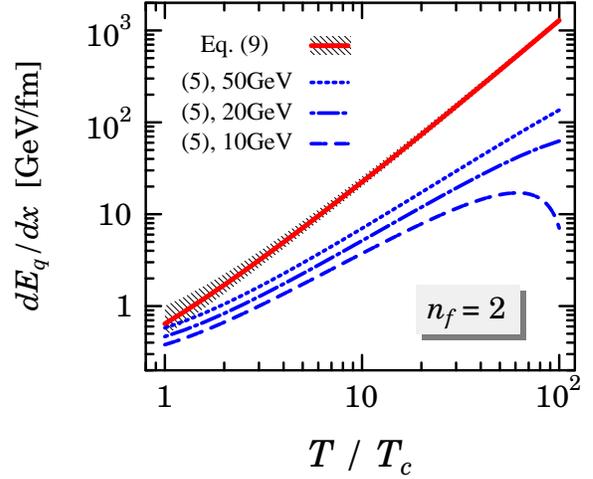}}
\caption{Light quark collisional energy loss: Eq.~(\ref{EQ.dEdx}) vs.\ the 
	prevalent expression (\ref{EQ.dEdx_Bjorken}) (which yields negative values
	for very large $T$) for representative jet energies. For details see text.}
\label{fig:dEdx}
\end{figure}

With this motivation, an extrapolation of Eq.~(\ref{EQ.dEdx}) as shown in Fig.~\ref{fig:dEdx} might be not too unreasonable.
Assumed here is $T_c = 160\,$MeV and $\mu^2 = [\frac12 ... 2] m_D^2$ to estimate the uncertainty due to the IR cut-off.
Shown for comparison are results from (\ref{EQ.dEdx_Bjorken}); here $\mu = m_D$ (likewise from (\ref{EQ.mD})) albeit $\alpha$ in the prefactor (unjustified, but as commonly presumed) fixed at the scale $Q_T = 2\pi T$.
It turns out that already near $T_c$, the estimates from formula (\ref{EQ.dEdx}) exceed those from (\ref{EQ.dEdx_Bjorken}), even for rather large values of $E$.

\begin{figure}
\resizebox{0.5\textwidth}{!}{\includegraphics{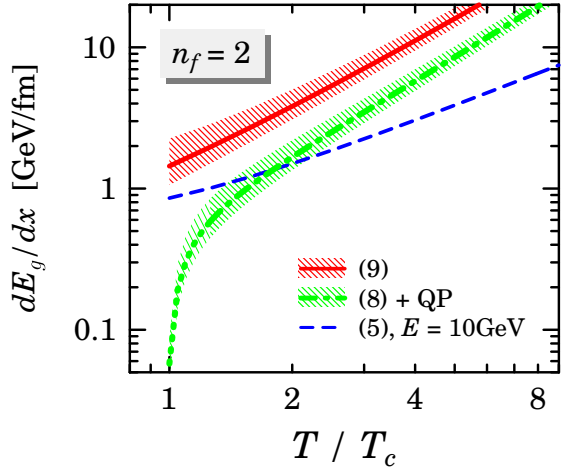}}
\caption{The diminished scatterer density near $T_c$ reduces the
	energy loss, as shown here (for the case of gluons) by the 
	dash-dotted curve. Below $\tilde T \approx 1.2T_c$, Eq.~(\ref{EQ.mD}) 
	over-estimates the actual Debye mass (as obtained in lattice QCD); hence 
	$dE/dx$ could be slightly larger than depicted by the dotted line.}
\label{fig:dEdx_extra}
\end{figure}

With regard to phenomenology it is instructive to take into consideration a principal effect of the strong interaction near $T_c$. From the distinct decrease of the QGP entropy established in lattice QCD calculations \cite{Karsch:2000ps}, one can, on rather general grounds, infer a similar behavior for the number densities $\rho_s$. In the framework of the quasiparticle model \cite{Peshier:1999ww}, the particle densities in Eq.~(\ref{EQ.dEdx_before_k-int}) read $d_s n_\pm(\sqrt{m_s^2(T)+k^2})$, where the effective coupling in the quasiparticle masses, $m_s^2 \propto \alpha_{\rm eff}(T) T^2$, is adjusted to lattice results for the entropy.
The effect on the energy loss, as exemplified in Fig.~\ref{fig:dEdx_extra}, is sizable: near $T_c$ -- despite the strong interaction -- the QGP could become transparent due to the reduced number of `active' degrees of freedom.
It should be noted, however, that the detailed behavior of the (collisional) energy loss near $T_c$ is speculative. For instance, a distinctly dropping Debye mass, as suggested by some lattice calculations, could (partly) counterbalance the diminishing effect of the dropping particle density.

\section{Conclusions}
Reverting, for a resum\'e, to a regime where perturbation theory can be expected viable, it has been demonstrated that taking into account the running of the strong coupling can influence the {\em structure} of results.
For the collisional energy loss of hard partons, the result changes by one order in the coupling compared to previous expressions derived with a fixed $\alpha$, cf.\ Eq.~(\ref{EQ.dEdx}) vs.\ (\ref{EQ.dEdx_Bjorken}).
While the increasing coupling at soft momenta leads to the parametric enhancement, asymptotic freedom, on the other hand, implies that the collisional energy loss becomes independent of the jet energy for $E \rightarrow \infty$. Consequently, the asymptotic behavior of the underlying interaction makes the energy loss {\em qualitatively} different in QCD and QED (where an analog of Eq.~(\ref{EQ.dEdx_Bjorken}) indeed holds).

Except very near $T_c$, Eq.~(\ref{EQ.dEdx}) suggests a larger collisional energy loss than previously estimated \cite{Bjorken:1982tu,Braaten:1991jj}. This finding are a facet of the `strongly coupled' QGP, which is characterized by large interaction rates. In fact, cross sections $\sigma = \int dt\, d\sigma/dt$ with {\em running} coupling are actually almost an order of magnitude larger than expected from the widely used expression $\sigma_{\alpha\, {\rm fix}} \sim \alpha^2(Q_T^2)/\mu^2 \sim 1$mb.
Thus, the present approach gives also a consistent explanation of the phenomenologically inferred large partonic cross sections ${\cal O}(10$mb) \cite{Molnar:2001ux,Peshier:2005pp} in the `sQGP'.

\bigskip

\noindent
{\bf Acknowledgments:} I would like to thank the organizers of {\em Hot Quarks 2006} for a distinguished and enlightening workshop.
My work was supported by BMBF and DFG.

\end{document}